\documentclass[a4paper,11pt]{scrartcl}
\usepackage[centertags]{amsmath}
\usepackage{amsfonts}
\usepackage{amssymb}
\usepackage{amsthm}
\usepackage{nicefrac}
\usepackage{titlesec}
\usepackage[ruled, section]{algorithm}
\usepackage{pifont}
\usepackage[ansinew]{inputenc}
\usepackage{graphicx}
\usepackage{natbib}
\usepackage{color}

\theoremstyle{plain}

\begin{document}

\title{Calculation of the mean duration and age of onset of a chronic disease and application to
dementia in Germany}

\author{Ralph Brinks\footnote{rbrinks@ddz.uni-duesseldorf.de}\\
Institute for Biometry and Epidemiology\\German Diabetes Center\\
Düsseldorf, Germany}

\date{}

\maketitle

\begin{abstract}
This paper descibes a new method of calculating the mean duration and mean age of onset of a chronic disease from 
incidence and mortality rates. It is based on an ordinary differential equation resulting from a simple 
compartment model. Applicability of the method is demonstrated in data about dementia in Germany. 
\end{abstract}


\section{Introduction}
This paper deals with analytical methods for calculating the mean age of onset and the
mean disease duration in chronic diseases. For this calculation so-called
IPM models are used. IPM stands for incidence, prevalence and mortality. 
First, the general IPM model is introduced. 
Then, formulas for the mean duration and the mean age of onset of the disease 
will be developed and applied to epidemiological data about dementia in Germany.

\begin{figure*}[ht]
\centerline{\includegraphics[keepaspectratio,
width=14cm]{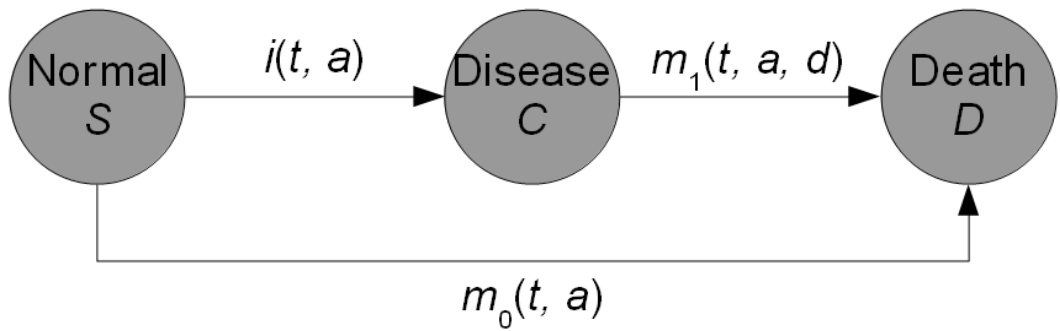}} \caption{Simple model 
of a chronic disease with three states. Persons in the state
\emph{Normal} are healthy with respect to the considered disease.
In the state \emph{Disease} they suffer from the disease. The
transition rates depend on the calendar time $t,$ on the age $a,$ and
in case of the disease-specific mortality $m_1$ also on the
disease's duration $d$.} \label{fig:3states}
\end{figure*}

Since we are interested in basic epidemiological parameters such as incidence, prevalence 
and mortality with respect to a disease, it has proven to be helpful to look at 
so-called state models (compartmental models). The model that will be used here consists of
the three states \emph{Normal}, \emph{Disease}, \emph{Death} 
and the transitions between the states. \emph{Normal} means non-diseased with respect to the disease
under consideration. The numbers of persons in the \emph{Normal} and 
\emph{Disease} state are denoted as $S$ (susceptible) and $C$ (cases), respectively.
The transition intensities (synonymously: rates) are called as shown in Figure \ref{fig:3states}: 
$i$ is the incidence rate, $m_0$ and $m_1$ are the mortality rates of the non-diseased 
and diseased persons, respectively. In general, the intensities depend on calendar time 
$t$, age $a$ and sometimes also on the duration $d$ of the disease.

Henceforth, we assume that the rates do not depend on calendar time and not 
on the duration of the disease. In this situation
\citeauthor{Mur94} considered a two-dimensional system of ordinary differential equations (ODEs), 
which expresses the change in the numbers of healthy and sick patients aged $a$ with 
the corresponding rates, \citeyearpar{Mur94}: 

\begin{equation}\label{eq:MurrayODE}
\begin{split}
    \frac{\mathrm{d} S}{\mathrm{d} a} &= - \left ( i(a) + m_0(a) \right ) \cdot S\\
    \frac{\mathrm{d} C}{\mathrm{d} a} &= i(a) \cdot S - m_1(a) \cdot C.\\
\end{split}
\end{equation}

Age plays here the role of temporal progression. 
In the paper \citep{Bri11} this two-dimensional system is used to derive a scalar 
ODE of Riccati type, which relates the change in the prevalence
at age $a$ to the rates $i$, $m_0$ and $m_1$:

\begin{equation}\label{eq:eq4}
\frac{\mathrm{d} p}{\mathrm{d} a} = (1-p) \cdot \left ( i - p \cdot \left (m_1 -
m_0 \right ) \right ).
\end{equation}

In general, this Ricatti ODE has no closed analytical solution, we have to be contented 
with numerical solutions. However, there are special cases (e.g. $m_0 \equiv m_1$) in which the ODE is linear, 
and then there are explicit expressions for the solution.

The benefits of such ODEs are obvious. For given age-specific incidence and mortality rates the
age profile of the prevalence can be obtained by solving the Ricatti ODE. Typically this is called
the forward or direct problem: we infer from the causes (the rates) to the effect (the prevalence). 
With smoothness constraints, the incidence and mortality rates determine the prevalence in an unambiguous way. 

In addition, for given prevalence and mortality rates the incidence can be dissolved.
This is the inverse problem -- we conclude from the effect to the cause. This allows, 
for example, cross-sectional studies being used for incidence estimates,
for what otherwise lengthy follow-up studies would be necessary. An example of such an application 
is shown in \citep{Bri11}.

\clearpage

\section{Mean disease duration and median age of onset}
In this terminology, formulas for the mean age $\overline{A}$ of onset  
and the average duration $\overline{D}$ of a chronic disease can be derived. 
After the preparatory work in the introduction this is now quite simple:
For the mean duration $\overline{D}$ the total person time spent in the disease state, 
i.e. the integral over $C(a)$, has to be 
divided by the total number of new cases. The total number of new cases is the integral over the product of 
incidence and number of susceptibles. Hence, it holds

\begin{equation}\label{eq:meanDuration}
\overline{D} = \frac{\int\limits_0^\omega C(a) \mathrm{d}a}{\int\limits_0^\omega i(a) \cdot S(a) \mathrm{d}a}
= \frac{\int\limits_0^\omega p(a) \cdot N(a) \mathrm{d}a}{\int\limits_0^\omega i(a) \cdot (1-p(a)) \cdot N(a) \mathrm{d}a},
\end{equation}
where
$N(a) := S(a) + C(a)$ is the total number of persons in the population with age $a$ and $\omega$ is an age that
exceeds the age of the oldest member of the population. 
In interpreting this expression as the first moment of a random variable $D$, the corresponding variance is
\begin{equation}\label{eq:varDuration}
\operatorname{Var}(D) = \frac{\int\limits_0^\omega \left ( C(a) - \overline{D} \right )^2 \mathrm{d}a}
                  {\int\limits_0^\omega i(a) \cdot S(a) \mathrm{d}a}
= \frac{\int\limits_0^\omega \left ( p(a) \cdot N(a) - \overline{D} \right )^2 \mathrm{d}a}
                  {\int\limits_0^\omega i(a) \cdot (1-p(a)) \cdot N(a) \mathrm{d}a}.
\end{equation}

\noindent For the mean age $\overline{A}$ of onset, the new cases are weighted by the corresponding age. 
\begin{equation}\label{eq:meanAge}
\overline{A} = \frac{\int\limits_0^\omega a \cdot i(a) \cdot S(a) \mathrm{d}a}{\int\limits_0^\omega i(a) \cdot S(a) \mathrm{d}a}
= \frac{\int\limits_0^\omega a \cdot i(a) \cdot \left ( 1-p(a) \right ) \cdot N(a) \mathrm{d}a}{\int\limits_0^\omega i(a) \cdot (1-p(a)) \cdot N(a) \mathrm{d}a},
\end{equation}
Correspondingly, the variance of the age of onset is
\begin{equation}\label{eq:varAge}
\operatorname{Var}(A) = \frac{\int\limits_0^\omega \left ( a - \overline{A} \right )^2 \cdot i(a) \cdot S(a) \mathrm{d}a}{\int\limits_0^\omega i(a) \cdot S(a) \mathrm{d}a}
= \frac{\int\limits_0^\omega \left ( a - \overline{A} \right )^2  \cdot i(a) \cdot \left ( 1-p(a) \right ) \cdot N(a) \mathrm{d}a}{\int\limits_0^\omega i(a) \cdot (1-p(a)) \cdot N(a) \mathrm{d}a}.
\end{equation}

Now we have these representations of $\overline{D}$ and $\overline{A}$. Since the age-specific prevalence 
$p$ is given by the ODE \eqref{eq:eq4}, $\overline{D}$ and $\overline{A}$ are completely determined by rates 
$i, m_0, m_1$ and $N$. It is remarkable, that $\overline{D}$ and $\overline{A}$ depend on the shape $N(a)$ of the
age pyramid. One might expect that these numbers are characteristics that just depend on the disease, that they are
disease inherent. However, easy considerations show that $\overline{D}$ and $\overline{A}$ have to be dependent
on the age-distribution of the population.

The function $N(a)$ is subject of demography. There are in the population models
where the numbers of people in the age groups 
can be represented analytically. The simplest example is the so-called stationary population, \citep{Pre82}, 
which can for example be obtained from the following conditions:

\begin{enumerate}
 \item Mortality rates do not depend on calendar time.
 \item The birth rate does not depend on calendar time.
 \item At each age level, the net migration is 0.
\end{enumerate}

For a stationary population, there are explicit representations for $N(a).$ However, real populations 
typically are non-stationary, which have to be managed differently. 
In Germany, there is an official population statistics from the Federal Statistical Office, 
which captures the age structure of the German population accurately. 
With $N(a)$ given for every age $a=0, \dots, 100+$ from the official
statistics, the integrals in Equations \eqref{eq:meanDuration} -- \eqref{eq:varAge} are replaced by sums.

\section{Application to dementia in Germany}
Our goal now is to apply the tools developed above to epidemiological data about dementia in Germany. 
The age-specific incidence is taken from \citep{Zie09}, which reports data for males and
females separately. The mortality $m$ in the German general population is taken from 
the current life table of the Federal Statistical Office \citeyearpar{Des11}. Reference year is 2010. 
The relative mortality of the patients is set constant to $R=2.4$, \citep{Rai10}, as in \citep{Bri11}.
In case the general mortality $m$ and the relative
mortality $R$ are given, the ODE \eqref{eq:eq4} changes its type and becomes Abelian, \citep[Tab. 1]{Bri11}:

\begin{equation}\label{eq:eq4Abel}
\frac{\mathrm{d} p}{\mathrm{d} a} = (1-p) \cdot \left \{ i - m \cdot \left( 1 - \left [ p \cdot \left ( R -
1 \right ) + 1 \right ]^{-1} \right ) \right \}.
\end{equation}

\begin{figure*}[ht]
\centerline{\includegraphics[keepaspectratio,
width=14cm]{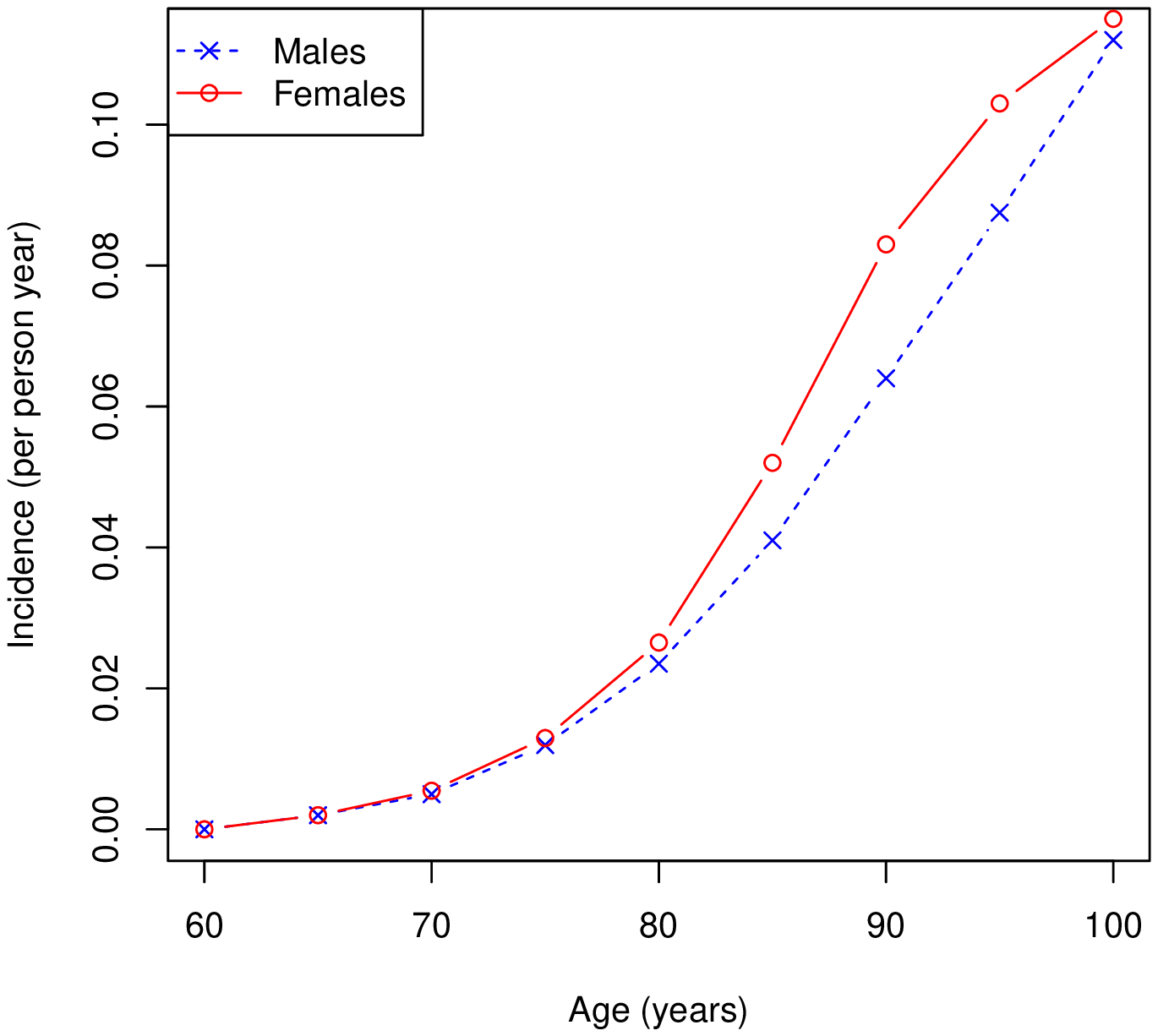}} \caption{Age- and sex-specific incidence of dementia in Germany, \citep[Tab. 3]{Zie09}.} 
\label{fig:inz}
\end{figure*}

Figure \ref{fig:inz} shows the age-specific incidence of dementia for males and females as reported in \citep{Zie09}. 
For this work, the data have interpolated affine-linearly using the middle of the age classes as knots.

\clearpage

\begin{figure*}[ht]
\centerline{\includegraphics[keepaspectratio,
width=14cm]{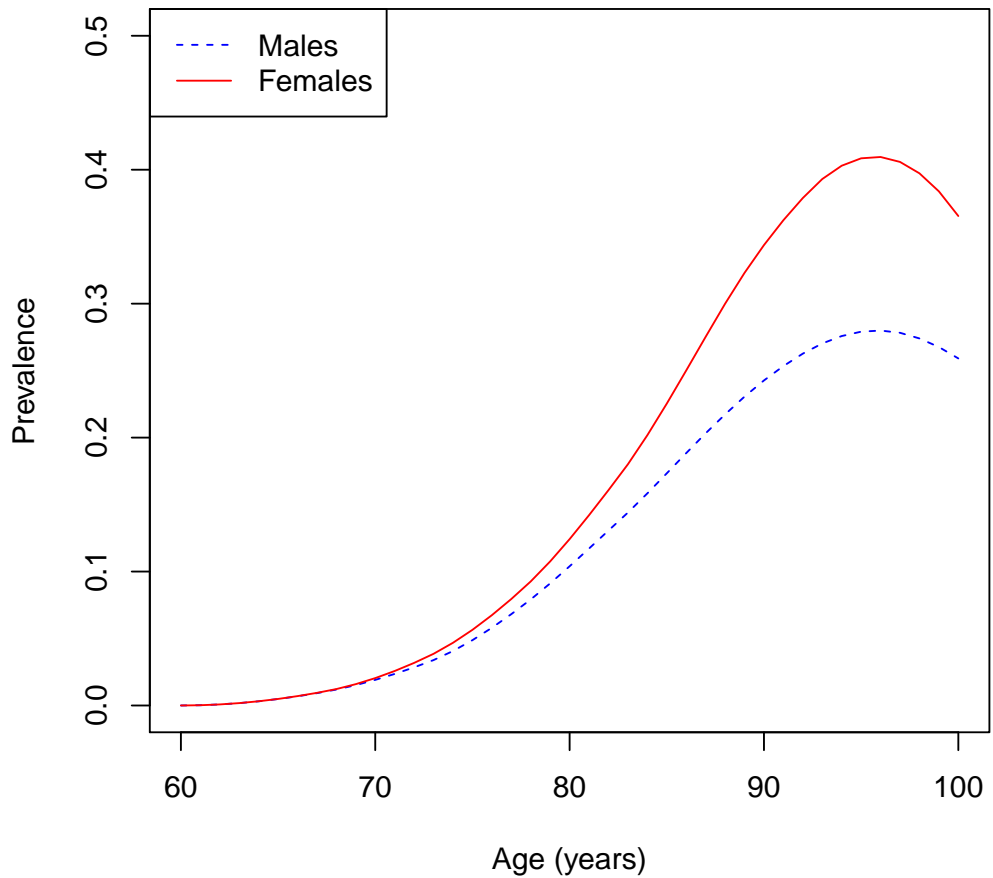}} \caption{Age- and sex-specific prevalence of dementia after integration of the
ODE \eqref{eq:eq4Abel}.} \label{fig:prev}
\end{figure*}

The age-specific prevalence is derived by integrating the Abelian ODE \eqref{eq:eq4Abel} with initial
condition $p(60) = 0$ via the classical Runge-Kutta method (of fourth order), cf. e.g. \citep{Dah74}. All
calculations are performed with the Software R (The R Foundation for Statistical Computing), version 2.12.0.
The result is shown in Figure \ref{fig:prev}.

Prevalence at age 60 years starts at 0, which is the initial condition. Until 70 years of age prevalence of
men and women are almost the same. In the early 70ies the curves start to diverge, which is likely to be an
effect of the different general mortality. Incidence rates in this age class are almost the same for men and women,
however general mortality $m$ for men is almost twice as high as for women. A higher general mortality 
$m_\text{male} > m_\text{female}$ in Eq. \eqref{eq:eq4Abel} leads to a lower gradient 
$\tfrac{\mathrm{d} p_\text{male}}{\mathrm{d} a} < \tfrac{\mathrm{d} p_\text{female}}{\mathrm{d} a}.$ It is striking
that both prevalence curves have a maximum at age $a^\star = 96$ years. At this age it holds 
$\tfrac{\mathrm{d} p}{\mathrm{d} a} = 0$, which implies $i(a^\star) = p(a^\star) \cdot m_0(a^\star) \cdot (R - 1)$.

\bigskip

\begin{figure*}[ht]
\centerline{\includegraphics[keepaspectratio,
width=14cm]{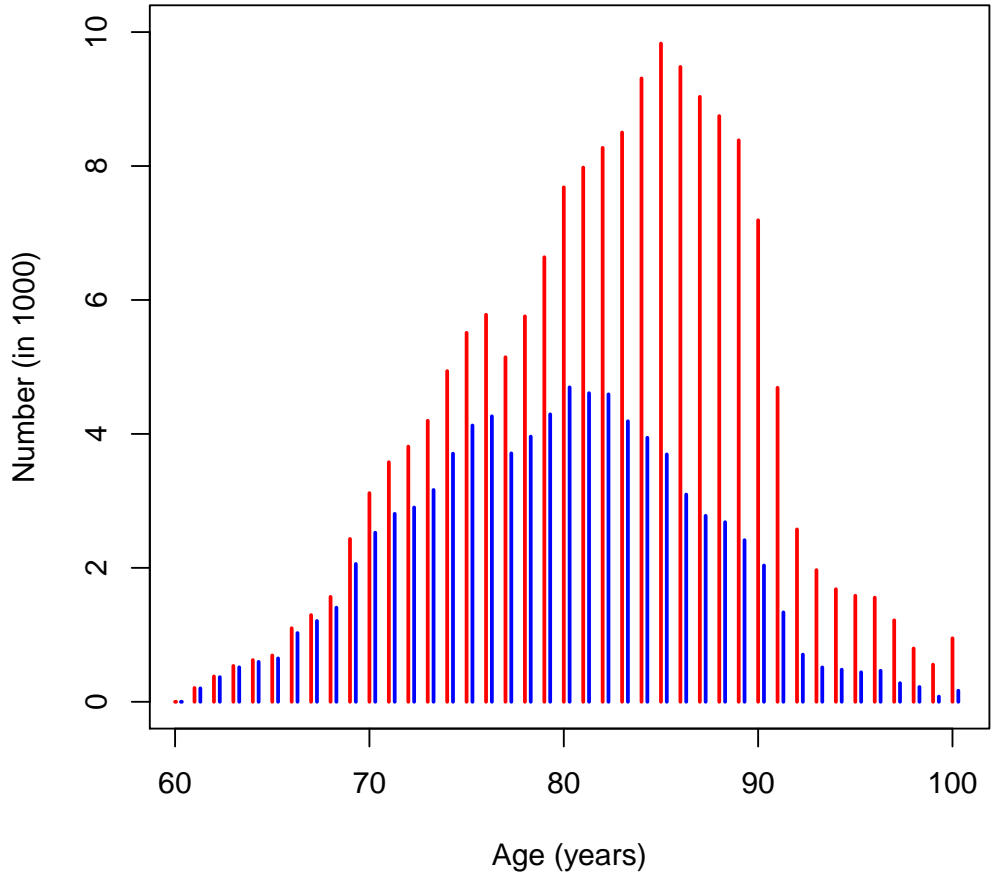}} \caption{Age distribution of the new cases (blue: males, red: females)
in 2010 based on the age pyramid $N(a)$ of Germany in the same year.} \label{fig:ageDistr}
\end{figure*}

The age distribution of the new cases $i(a) \cdot (1-p(a)) \cdot N(a)$ for each age $a = 50, \dots, 100+$ in 2010
is shown in Figure \ref{fig:ageDistr}. 

The modus of the new cases is 80 and 85 years in men and women,
respectively. Obviously, women are far more affected than men. The discontinuities stem from the 
discontinuous structure of the age pyramid.

\clearpage

The corresponding mean age $\overline{A}$ of onset and the mean duration $\overline{D}$ are presented in
Table \ref{tab:results}.
\begin{table}[ht]
 \centering
 \def\~{\hphantom{0}}
 \begin{minipage}{150mm}
  \caption{Mean age of onset and duration of dementia in Germany in 2010.}\label{tab:results}
  \begin{tabular*}{\textwidth}{@{}l@{\extracolsep{\fill}}l@{\extracolsep{\fill}}l@{\extracolsep{\fill}}}
  \hline \hline
       & Mean age $\overline{A}$  &  Mean duration $\overline{D}$ \\ 
	   & (in years)               &  (in years) \\ \hline
Males  &  79.3 & 4.6 \\
Females&  82.1 & 5.4 \\
\end{tabular*}
\end{minipage}
\vspace*{-6pt}
\end{table}


\section{Summary}
In this work a simple IPM model has been used to study the mean age of onset and mean duration
of a chronic disease. These numbers depend on the incidence of the disease, the mortality of both
diseased and non-diseased persons and also on the age pyramid of the population under consideration. 
The age-specific prevalence inherent in both numbers was obtained as the solution of a new ODE, 
which has been derived in the predecessor paper \citep{Bri11}. As a practical example, the calculations
were applied to data about dementia in Germany.
The mean age of onset of dementia is about 79 and 82 years with mean duration 4.6 and 5.4 years 
in men and women, respectively. Due to the different life
expectancies of men and women in Germany, the differences are not surprising. However, it is striking 
how big the difference in the nubmers of the cases of dementia are. Figure \ref{fig:ageDistr} shows the differences
in each of the age groups. In the year 2010 a total of about 85000 men and 168000 women aged 65 years and above 
become diseased.
This is in the same order as the 78300 and 166000 males and females reported in \citep[Tab. 4]{Zie09} for the year 2007.
The reasons for the about doubled number of women compared to men are on the one hand the higher incidence of dementia 
in females, which leads to a far higher prevalence (see Fig. \ref{fig:prev}), and on the other hand the higher number 
of females aged 65+. For comparison, in Germany in 2010 there are 7.2 and 9.6 
million men and women aged 65 years and above, respectively.

\bigskip

The methods described here are advantagous to predict characteristics of the persons with dementia. The age of
onset might be important for comorbidities (such as diabetes) and associated late complications. The estimated
number of diseased persons and disease durations are highly relevant for health services allocation planning.
The methods allow predictions on regional levels, too. The age distributions of the states and communities 
in Germany differ quite substantially, which means that the functions $N(a)$ in the states and communities 
are different. Hence the associated mean ages of onset and mean durations are different.

This study has some weaknesses. First, the method described needs the incidence and mortality rates
being independent from calendar time and disease duration. Both assumptions are hardly fulfilled. Hence,
the calculations shown here are just an approximation. Second,
the incidences are based on claims data of the statutory health insurance (SHI) from 2002. Beside the fact
that the data are old compared to the reference year 2010, SHI data tend to overestimate incidence, \citep{Abb11}.
Third, the ages equal to and above 100 are summarized in an age class 100+. The reason is the age pyramid of the Federal 
Statistical Office, which does not stratify ages beyond 100 years.

Another point is worth being mentioned: the examinations in this paper predict a peak in the prevalence
of dementia in the second half of the ninth decade of life for men and women. After the peak, prevalence is
decreasing again. If sufficient data is available for this age group, this hypothesis can be tested.

\appendix

\label{lastpage}

\end{document}